\documentclass[10pt,conference]{IEEEtran} 
\IEEEoverridecommandlockouts

\usepackage[flushleft]{threeparttable}
\usepackage{color}
\usepackage{multirow}
\usepackage{listings}
\usepackage{bm}
\usepackage[colorlinks=true]{hyperref}
\usepackage{amsmath}
\usepackage{booktabs}
\usepackage{balance}
\usepackage[most]{tcolorbox}

\newcommand{\appname}{{\textsc{JointCom}}}

\usepackage{enumitem}
\setlist[itemize]{left=0em}

\usepackage{pgfplots}
\pgfplotsset{compat=1.17}

\begin{document}
\title{Improving Retrieval-Augmented Code Comment Generation by Retrieving for Generation}

\author{
\IEEEauthorblockN{Hanzhen Lu\IEEEauthorrefmark{2}, Zhongxin Liu\IEEEauthorrefmark{1}\IEEEauthorrefmark{2}\IEEEauthorrefmark{3}\thanks{\IEEEauthorrefmark{1}Corresponding author.}} 
\IEEEauthorblockA{\IEEEauthorrefmark{2}The State Key Laboratory of Blockchain and Data Security, Zhejiang University, China\\
\IEEEauthorrefmark{3}Hangzhou High-Tech Zone (Binjiang) Institute of Blockchain and Data Security, China\\
\{luhanzhen, liu\_zx\}@zju.edu.cn
}
}

\maketitle

\begin{abstract}
    Code comment generation aims to generate high-quality comments from source code automatically and has been studied for years.
    Recent studies proposed to integrate information retrieval techniques with neural generation models to tackle this problem, i.e., Retrieval-Augmented Comment Generation (RACG) approaches, and achieved state-of-the-art results.
    Generally, RACG approaches use a retriever to retrieve a code-comment pair from a retrieval base as an exemplar, combine the exemplar with the input code snippet, and feed the combined text to a generator (usually a sequence-to-sequence model) to generate the comment.
    However, the retrievers in previous work are built independently of their generators.
    This results in that the retrieved exemplars are not necessarily the most useful ones for generating comments, limiting the performance of existing approaches.
    To address this limitation, we propose a novel training strategy to enable the retriever to learn from the feedback of the generator and retrieve exemplars for generation.
    Specifically, during training, we use the retriever to retrieve the top-k exemplars and calculate their retrieval scores, and use the generator to calculate a generation loss for the sample based on each exemplar.
    By aligning high-score exemplars retrieved by the retriever with low-loss exemplars observed by the generator, the retriever can learn to retrieve exemplars that can best improve the quality of the generated comments.
    Based on this strategy, we propose a novel RACG approach named \appname\  and evaluate it on two real-world datasets, JCSD and PCSD.
    The experimental results demonstrate that our approach surpasses the state-of-the-art baselines by 7.3\% to 30.0\% in terms of five metrics on the two datasets.
    We also conduct a human evaluation to compare \appname\  with the best-performing baselines.
    The results indicate that \appname\  outperforms the baselines, producing comments that are more natural, informative, and useful.
\end{abstract}

\section{Introduction}\label{sec:introduction}
Nowadays, the scale of software has significantly increased, causing programmers to spend a significant amount of time understanding and maintaining source code~\cite{ko2006exploratory, xia2017measuring}.
Comments play a crucial role in hiding complex details of code and making code understanding easier~\cite{de2005study}.
However, the manual creation and upkeep of code comments during code changes are both time-consuming and labor-intensive~\cite{de2005study, kajko2005survey}.
As a result, researchers proposed to automatically generate high-quality comments for code.

Considering that code reuse is prevalent during software development~\cite{kim2005empirical, kamiya2002ccfinder}, some prior work~\cite{wong2013autocomment,wong2015clocom} employed information retrieval (IR) techniques to retrieve similar code snippets from a retrieval base and reuse the corresponding comments as the generated comments.
A retrieval base can be existing code repositories, Q\&A sites, a collected training set, etc.
Subsequently, with the development of deep learning, many studies~\cite{CodeNN, hu2018deep, TLCodeSum} proposed to regard comment generation as a language translation task and employed neural network models, usually sequence-to-sequence (seq2seq) models, to generate comments from source code.
Recently, researchers~\cite{arthur2016incorporating,zhang2018guiding} noticed that neural methods generally prefer high-frequency tokens and seldom generate low-frequency tokens, which can be complemented by IR-based methods. 
Thus, they proposed to combine IR-based methods and neural network-based methods for comment generation and achieve state-of-the-art performance~\cite{re2com, Rencos, editsum, decom}.
We refer to these approaches as Retrieval-Augmented Comment Generation (RACG) approaches.
Generally, given a code snippet, RACG approaches use a retriever to retrieve a code-comment pair from the retrieval base as the exemplar, combine both the exemplar with the input code snippet, and feed the combined text to a neural model to generate the comment.
A good exemplar not only enhances the ability of the neural model to generate low-frequency words~\cite{re2com} but also serves as a template that provides a lot of reusable patterned words for comment generation~\cite{editsum}.

\begin{table}[!t]
    \small
    \centering
    \caption{An example of code comment generation}
    \label{tab:Introduction_Example}
    \vspace{-0.25cm}
    \begin{threeparttable}
    \addtolength\tabcolsep{2.8pt}
    \begin{tabular}{|p{0.47\textwidth}@{\hskip3pt}|}
         \toprule
         \textbf{Code 1}:\\
         \vspace{-0.5cm}
         \begin{lstlisting}[language=Python]
'change private-browsing config to true and emit signal'
def test_cache_config_enable_private_browsing(config_stub, tmpdir): 
    config_stub.data = {'storage': {'cache-size': 1024}, 'general': {'private-browsing': False}} 
    disk_cache = cache.DiskCache(str(tmpdir)) 
    assert (disk_cache.cacheSize() == 0) 
    preload_cache(disk_cache) 
    assert (disk_cache.cacheSize() > 0) 
    config_stub.set('general', 'private-browsing', True) 
    assert (disk_cache.cacheSize() == 0)
         \end{lstlisting}\vspace{-0.5cm}\\
         \hline
         \textbf{Code 2}:\\
         \vspace{-0.5cm}
         \begin{lstlisting}[language=Python]
'test if cache is empty after clearing it'
def test_cache_clear_activated(config_stub, tmpdir): 
    config_stub.data = {'storage': {'cache-size': 1024}, 'general': {'private-browsing': False}} 
    disk_cache = cache.DiskCache(str(tmpdir)) 
    assert (disk_cache.cacheSize() == 0) 
    preload_cache(disk_cache) 
    assert (disk_cache.cacheSize() != 0) 
    disk_cache.clear() 
    assert (disk_cache.cacheSize() == 0)
         \end{lstlisting}\vspace{-0.5cm}\\
         \hline
         \textbf{Code 3}:\\
         \vspace{-0.5cm}
         \begin{lstlisting}[language=Python]
'change private-browsing config to false and emit signal'
def test_cache_config_disable_private_browsing(config_stub, tmpdir): 
    config_stub.data = {'storage': {'cache-size': 1024}, 'general': {'private-browsing': True}} 
    url = 'http://qutebrowser.org' 
    metadata = QNetworkCacheMetaData() 
    metadata.setUrl(QUrl(url)) 
    assert metadata.isValid() 
    disk_cache = cache.DiskCache(str(tmpdir)) 
    assert (disk_cache.prepare(metadata) is None) 
    config_stub.set('general', 'private-browsing', False) 
    content = 'cute' 
    preload_cache(disk_cache, url, content) 
    assert (disk_cache.data(QUrl(url)).readAll() == content) 
         \end{lstlisting}\vspace{-0.5cm}\\
         \hline
         \textbf{Prediction for Code 1}\\
         \textbf{Using Code 2 as the exemplar}: test if cache is enabled after clearing it .\\
         \textbf{Using Code 3 as the exemplar}: change private-browsing config to true and emit signal .\\
         \bottomrule
    \end{tabular}
    \end{threeparttable}
    \vspace{-0.5cm}
\end{table}

Two types of methods are used by existing RACG approaches to retrieve exemplars, i.e., traditional IR methods~\cite{re2com, editsum, decom} and dense retrievers~\cite{Rencos}.
BM25 is the most widely used IR method, which calculates the relevance score between two code snippets based on lexical matching.
Instead, dense retrievers calculate such relevance scores based on a trained neural encoder.
For example, Rencos proposed by Zhang et al.~\cite{Rencos} first trained a seq2seq model for comment generation and then calculated the relevance score between two code snippets based on their embeddings produced by the encoder of the trained model.
Although existing RACG approaches are shown to be effective, their retrievers and generators are built separately and the retrievers do not know which exemplars can benefit their generators most.
Consequently, the exemplars retrieved by these retrievers are not necessarily the most useful ones for generation, limiting the performance of existing RACG approaches.
For example, considering the code snippets in Table~\ref{tab:Introduction_Example}, we trained a dense retriever and a retrieve-augmented generator following Rencos using the training set of the Python Code Summarization Dataset (PCSD)~\cite{barone2017parallel} and used both BM25 and the dense retriever to retrieve exemplars from the training set for Code 1.
Both retrievers regard Code 2 as more relevant than Code 3 to Code 1.
However, we found that using Code 3 as the exemplar can help the generator generate better comments, as shown in Table~\ref{tab:Introduction_Example}.

This example inspires us by demonstrating that a retriever capable of selecting exemplars better suited to the generator can enhance RACG approaches.
Based on this observation, we propose to jointly train the retriever and the generator so that the retriever can be guided by the feedback of the generator and retrieve the exemplar that is most useful for generation.
However, it is non-trivial to enable joint training.
Because the interaction between the generator and the retriever only involves discrete signals, i.e., the selected exemplar.
We cannot simply connect the retriever to the generator and calculate the gradients of the retriever based on the loss of the generator through backpropagation.
To enable the synergism of the retriever and the generator, we propose a joint training strategy.
Given a training sample, instead of only retrieving one exemplar for generation, we retrieve the top-k exemplars and calculate their retrieval scores using the retriever. 
The loss of generating the ground truth based on each exemplar is calculated using the generator. 
Then, we construct the weighted sum of the generation losses of these top-k exemplars with the weights being their retrieval scores. 
Finally, we optimize this loss to obtain the gradients of both modules and enable joint training. 
This training strategy brings several benefits. 
First, this weighted loss can guide the retriever to rank the exemplars based on their helpfulness in guiding the generator to generate ground truth. 
For example, if the generation loss of an exemplar is low, the retriever will learn to assign more weights to this loss to reduce the final weighted loss. 
Second, this weighted loss can guide the generator to pay more attention to the most helpful exemplar due to its maximum weight. 
Thus, we hypothesize that this training strategy can make the trained retriever and generator more helpful to each other.

Based on this training strategy, we propose a RACG approach named \appname.
Specifically, during training, the retriever first leverages a code encoder to compute the embedding of each code snippet in the retrieval base. 
Next, for each training sample, the retriever retrieves the top-k exemplars from the retrieval base.
Then, for each retrieved exemplar, its code and comment are concatenated with the input code snippet, and the concatenated text is fed to a seq2seq model to calculate the corresponding generation loss.
Based on the training strategy, a weighted loss is then calculated over all exemplars.
The retriever and the generator are optimized based on this loss through backpropagation.
The procedure of inference is similar to that of training, but we only retrieve the top-1 exemplar and regard the comment generated based on this exemplar as output.

Considering the impressive effectiveness of pre-trained code models in understanding natural language and programming languages~\cite{guo2022unixcoder, feng2020codebert, wang2021codet5}, we implement \appname\  based on CodeT5~\cite{wang2021codet5}, a widely-used pre-trained seq2seq model for code.
To evaluate our approach, we conduct extensive experiments on two widely used datasets, JCSD~\cite{hu2018summarizing} and PCSD~\cite{barone2017parallel}, with five metrics including Corpus-level BLUE, Sentence-level BLEU, METEOR, ROUGLE-L, and CIDEr. 
Experimental results show that \appname\  outperforms the state-of-the-art baselines by 7.6\% to 28.4\% in terms of the five metrics on JCSD.
On PCSD, \appname\  improves the best-performing baselines in terms of the five metrics by 9.6\% to 30.0\%.
Our component analysis shows that our training strategy substantially contributes to the effectiveness of \appname, and both the retriever and the generator are beneficial for \appname.
We also conduct a human evaluation to assess the generated comments based on three aspects: naturalness, informativeness, and usefulness. 
The results demonstrate that \appname\  produces comments that are more natural, informative, and useful than the baselines.

The contributions of our work are shown as follows:
\begin{itemize}
    \item To the best of our knowledge, this is the first work that proposes to improve retrieval-augmented comment generation by jointly training retrievers and generators.
    \item We propose a training strategy to enable joint training, making the retriever and the generator more helpful to each other. Based on this strategy, we propose and implement a new retrieval-augmented comment generation approach \appname.
    \item We conduct both automatic and human evaluations to evaluate our approach and the results show that \appname\  outperforms state-of-the-art approaches by substantial margins.
\end{itemize}

In the subsequent sections of this paper, we provide further details. Section \ref{sec:Preliminaries} introduces some background knowledge. Section \ref{sec:Approach} elaborates on our approach. Section \ref{sec:ExperimentalSetup} details the experimental setup. Section \ref{sec:Results} presents our results and analysis. Section \ref{sec:HumanEvaluation} introduces the methods and results of human evaluation. Section \ref{sec:Discussion} discusses specific examples, the feasibility of \appname\  on LLM, and threats to validity. Section \ref{sec:RelatedWork} reviews related work. Section \ref{sec:Conclusion} concludes our work and points out future directions.

\section{Preliminaries}\label{sec:Preliminaries}
Given the remarkable performance of pre-trained models in natural language processing (NLP)~\cite{devlin2018bert, lewis2019bart}, some researchers have sought to replicate such success in code intelligence tasks~\cite{feng2020codebert, guo2022unixcoder}.
They design some unsupervised learning tasks and pre-train deep neural networks in vast programming language (PL) and natural language (NL) corpora to learn useful patterns.
Existing pre-trained models can be categorized into encoder-only, decoder-only, and encoder-decoder models.
Encoder-only models, such as CodeBERT~\cite{feng2020codebert}, process the entire input to create contextually rich embeddings or representations and are suitable for understanding tasks.
Decoder-only models, such as CodeGPT~\cite{lu2021codexglue}, generate one token at a time based on past outputs and possibly additional input.
While effective for tasks such as code completion, their unidirectional design is less suited for comprehension tasks.
Encoder-decoder models, such as CodeT5~\cite{wang2021codet5}, have an encoder that reads and encodes the entire input into a dense representation, and a decoder that takes this representation to generate the output sequentially.
Thus they are suitable for tasks that involve transforming an input into an output.
Given that the comment generation task involves both understanding PL and generating NL and considering that most state-of-the-art models in this domain utilize the encoder-decoder framework, we construct our approach based on the pre-trained encoder-decoder models for code.

\section{Approach}\label{sec:Approach}
\begin{figure*}[th]
    \vspace{-0.5cm}
    \centering
    \includegraphics[width=0.9\linewidth,trim=0 120 0 0, clip]{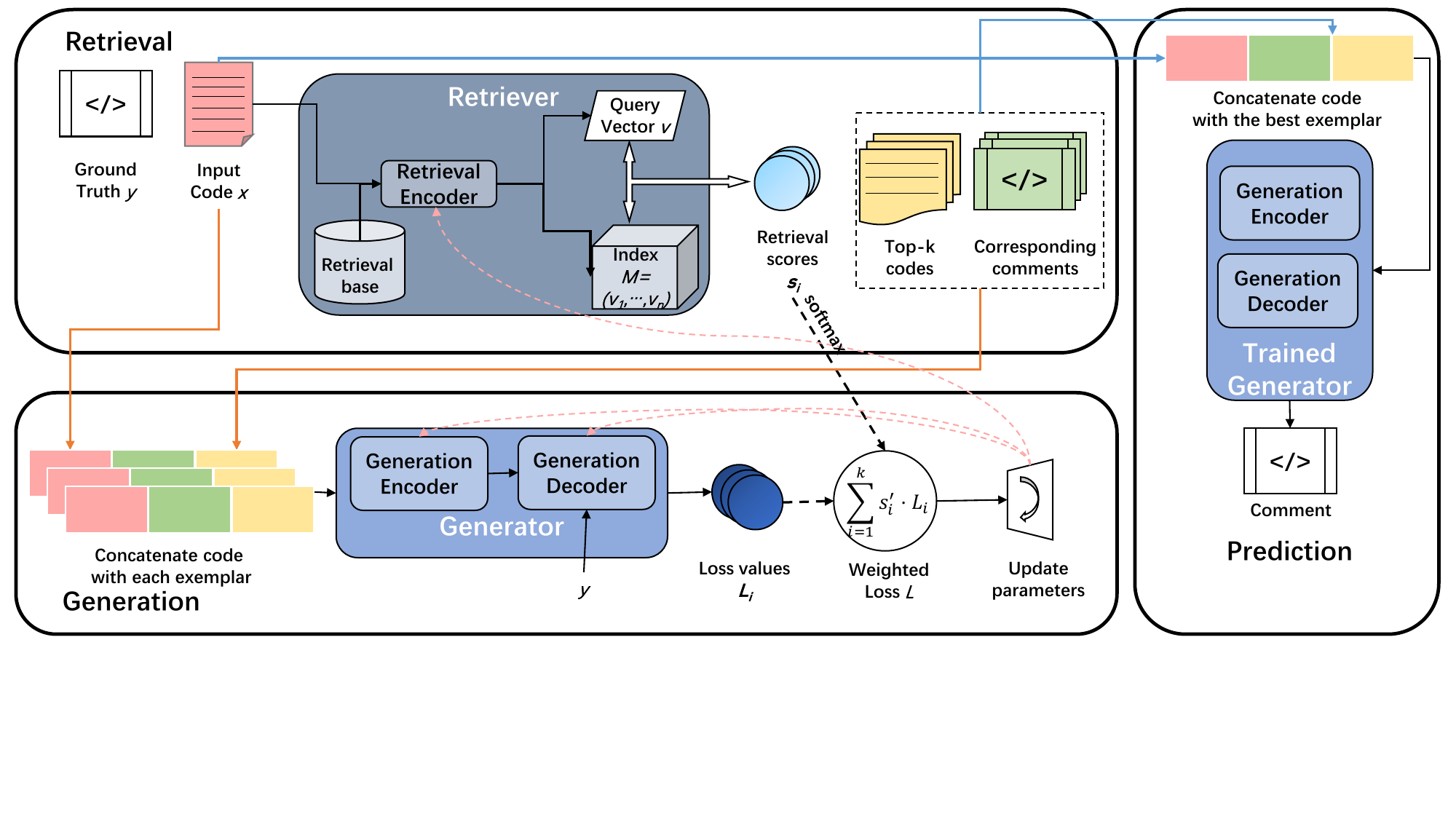}
    \caption{The overall framework of \appname}
    \label{fig:architecture}
    \vspace{-0.5cm}
\end{figure*}

Code comment generation aims to automatically generate high-quality comments for code.
This task can be formalized as finding the mapping $f$ from the input code snippet $x$ to its comment $y$, i.e., $y=f(x)$.
Deep-learning-based approaches tackle this task by training a neural network to model the probability $p(y|x)$.
RACG approaches divide this task into two stages, i.e., retrieval and generation.
In the retrieval stage, existing RACG approaches use a retriever to retrieve a code-comment pair (namely exemplar) $r^{*}$ from a retrieval base $R$.
In the generation stage, they train a neural generator to model the probability $p(y|x, r^{*})$.
As we discussed in Section~\ref{sec:introduction}, their limitation is that the retriever is built independently of the generator, and thus does not necessarily retrieve the exemplars that are the most helpful for generation.
To address this limitation, we propose an approach named \appname\  that jointly trains the retriever and the generator for better comment generation.

\subsection{Overall Framework}
Figure~\ref{fig:architecture} shows the overall framework of \appname.
Similar to existing RACG approaches, \appname\  contains two modules, i.e., the retriever and the generator.
Before training or prediction, the retriever leverages a neural encoder to encode the code snippets in the retrieval base into embeddings, namely key vectors, which are stacked into the search index.
During training, given a training sample, the retriever also uses the encoder to convert the code snippet into an embedding, namely a query vector.
The retrieval score of each exemplar in the retrieval base is calculated based on the query and key vectors, and the top-k exemplars are retrieved based on such scores.
The generator concatenates the input code snippet and each retrieved exemplar, including its code snippet and comment, as the generation input, and calculates the generation loss for this generation input. 
Then, \appname\  jointly optimizes the retriever and the generator based on our training strategy (described in Section~\ref{sec:approach:training}).
During prediction, only the exemplar with the highest retrieval score is retrieved for generation.

\subsection{Retrieval}\label{sec:approach:retrieval}
Given a code snippet, the retriever is responsible for retrieving its exemplars from the retrieval base for the generator.
As shown by the example in Table~\ref{tab:Introduction_Example}, for RACG approaches, the quality of the retrieved exemplars is important for generating high-quality comments.
Most previous work used a traditional IR technique named BM25 for retrieval.
However, BM25 only considers word overlaps, ignoring semantic similarities.
Also, it is not a parametric model and thus is not trainable.
To enable the joint training of the retriever and the generator, we follow Zhang et al.~\cite{zhang2020retrieval} and use a dense retriever.

The main component of our retriever is a neural encoder, e.g., a Transformer encoder, which encodes a code snippet into an embedding (feature vector).
Specifically, given a code snippet $x$, the encoder first tokenizes $x$ into a sequence of tokens and prepends a special token [CLS] to them to aggregate the information of all tokens. 
Then, it calculates the contextual embedding of each token using the neural network, e.g., the Transformer, and regards the contextual embedding of [CLS] as the embedding $v$ of $x$:
\begin{align}
    v = \text{encoder}( x )
\end{align}
where $v \in \mathcal{R}^{d}$ and $d$ is the hidden size of the encoder.
To expedite model convergence, we employ $L2$ normalization on the embedding, and each item in the embedding is modified as follows:
\begin{align}
    v_i' = \frac{v_i}{\sqrt{\sum_{j=1}^{d}v_j^2}}
\end{align}

Before training or prediction, the neural encoder calculates the embeddings of the code snippets in the retrieval base and stacks these embeddings into a matrix $M \in \mathcal{R}^{n\times d}$, where n is the size of the retrieval base.
$M$ is used as the search index.
Given a code snippet $x$, to retrieve its exemplars, the retriever calculates its embedding $v$ using the neural encoder and multiplies the embedding with $M$ to calculate a score vector $(s_1, s_2, \cdots, s_n)$, where $s_i$ refers to the retrieval score between the $i_{th}$ sample and the input code snippet.
\begin{align}
    (s_1, s_2, \cdots, s_n)^T &= M \cdot v^T
\end{align}
Due to our utilization of $L2$ normalization, the retrieval scores effectively represent cosine similarities. 
The retriever employs these scores to retrieve exemplars during both training and prediction phases.

\subsection{Generation}\label{sec:approach:generator}
The generator is responsible for generating a comment based on the input code snippet and an exemplar retrieved by the retriever.
Given a code snippet $x$ and a retrieved exemplar $r_i$, the generator first appends the comment $r^{\text{com}}_i$ and the code snippet $r^{\text{code}}_i$ of $r_i$ to construct the generation input $x'_i$.
To separate the three parts, we insert a \verb|\n| between each pair.
We also insert a \verb|#| at the beginning of the comment to signify that it is natural language.
The construction is shown as follows:
\begin{align}
    x_i' &= x \cdot \backslash \text{n\#} \cdot r^{\text{com}}_{i} \cdot \backslash \text{n} \cdot r^{\text{code}}_{i}
\end{align}
where $\cdot$ means concatenation.
Then, the generator leverages a seq2seq model, which typically consists of an encoder and a decoder, to generate the comment.
It first tokenizes the input into tokens using the tokenizer of the model and feeds these tokens into the encoder to obtain the contextual embedding of each token.
Then, these contextual embeddings are fed into the decoder to generate tokens one by one.
Specifically, at time step $t$, the last hidden state $h_t$ of the decoder is transformed into the probability distribution of the next token through a fully connected layer and the softmax function:
\begin{align}
    p(y_{t+1} | y_1, y_2, \cdots, y_t, x'_i) = \text{softmax}(h_tW + b)
\end{align}
where $y_{t}$ denotes the $t_{th}$ token on the ground truth comment $y$ and $W$ is a trainable matrix.
During training, we calculate the generation loss $\mathcal{L}_i$ for each $x'_i$ based on the cross entropy loss function:
\begin{align}
    \mathcal{L}_i=-\log p(y|x, r_i) = -\sum_{t=1}^{\left| y \right|} \log{p(y_t | y_{<t}, x'_i)} 
\end{align}
During prediction, the token with the highest probability is appended to the output sequence, and the token is fed back into the model to produce a prediction for the next token until a special token \verb|</s>|, which marks the end of a comment, is generated.

\subsection{Training}~\label{sec:approach:training}
Our approach aims to address the limitation of existing RACG approaches by using the preference of the generator to guide the retriever.
To achieve this goal, we need to assess the usefulness of each exemplar for the generator.
As described in Section~\ref{sec:approach:generator}, given a code snippet $x$ and a retrieved exemplar $r_i$, we construct a generation input $x_i'$ and calculate its generation loss $\mathcal{L}_i$.
The smaller $\mathcal{L}_i$ is, the more likely the ground truth can be generated based on $x_i'$.
Therefore, given two exemplars $r_i$ and $r_j$ of $x$, if the generation loss of $x_i'$ is smaller than that of $x_j'$, $r_i$ can be regarded as more useful than $r_j$ and should be assigned a higher retrieval score by the retriever.
Based on this observation and to optimize the retriever based on the preference of the generator, we propose the following loss function:
\begin{align}
    (s_1',s_2',\cdots,s_n')&=softmax(s_1,s_2,\cdots,s_n) \\
    \label{eq:prob}
    \mathcal{L} &= \sum_{i=1}^{n} \mathcal{L}_i \cdot s_i'
\end{align}
Using the softmax function, we first normalize the retrieval scores of all exemplars in the retrieval base, i.e., $s_1, s_2, \cdots, s_n$.
Next, we use the generator to generate a comment based on each exemplar $r_i$ and obtain the generation loss $L_i$.
Then, we define the Loss $L$ of \appname\  as the weighted sum of all the generation losses, and the normalized retrieval score $s_i'$ of $r_i$ is used as the weight of $L_i$.
This loss function enables us to jointly train both the retriever and the generator, as $L_i$ and $s_i'$ can guide the generator and the retriever to optimize.
On one hand, since we normalize the retrieval scores, the retriever will learn to assign a higher retrieval score to the exemplar with a lower generation loss to reduce the final loss $L$.
In this way, the retriever can learn to predict the helpfulness of each exemplar for comment generation. 
On the other hand, compared to existing RACG approaches where the generator is only trained on one exemplar, \appname's generator is trained with a broader range of instances, where the retrieved exemplars may be useful or useless.
This may increase the robustness of the generator and make it able to handle exemplars of different qualities during prediction.

However, calculating $\mathcal{L}$ based on Equation~\ref{eq:prob} requires applying softmax over all the $s_i$ and calculating the generation loss $L_i$ for each sample in the retrieval base, which can be very time-consuming considering a retrieval base typically contains a large number of samples.
To tackle this problem, we simplify $\mathcal{L}$ by only considering the k exemplars with the highest score, i.e., the top-k exemplars, as follows:
\begin{align}
    \label{eq:topk_prob}
    (s_{r_1}',\cdots,s_{r_k}')&=softmax_{r_i \in \textsc{top-k}}(s_{r_1},\cdots,s_{r_k}) \\
    \label{eq:Loss}
    \mathcal{L} &= \sum_{r_i \in \textsc{top-k}}\mathcal{L}_{r_i} \cdot s_{r_i}'
\end{align}
This simplification is reasonable because it is usually the case that only a few exemplars in the retrieval base can help generate the comment of a given code snippet.
Note that if the retrieved top-k exemplars are of low quality for all training samples, it would be difficult for our retriever to learn well. 
To avoid this extreme case, our approach initializes the neural encoder in the retriever using a high-quality pre-trained code encoder.

During training, after the retriever is updated, the embeddings calculated by this retriever should be updated too. 
However, updating the whole search index $M$ after each batch is time-consuming due to the large number of samples in the retrieval base.
To speed up training, we only re-calculate and update the embeddings of the retrieved top-k exemplars after updating the parameters.
To limit the deviation, the retriever updates $M$ after each epoch.
Following prior work~\cite{decom,re2com}, we directly use the training set as the retrieval base.
When retrieving exemplars for a training sample, the retriever will remove the training sample itself from the retrieval base.

\subsection{Prediction}
During prediction, our target is to generate a high-quality comment.
Therefore, given a code snippet, we only retrieve the exemplar with the highest retrieval score using the retriever.
Then we leverage the generator to generate the comment based on the code snippet and the exemplar.
Following prior work~\cite{decom}, beam search~\cite{wiseman2016sequence} is also employed to track the globally optimal result in the prediction stage.

\section{experimental Setup}\label{sec:ExperimentalSetup}
This section describes the datasets, the evaluation metrics we use to evaluate our approach, and the implementation details of our approach.

\subsection{Dataset}
\begin{table}[!t]
    \vspace{-0.3cm}
    \centering
    \caption{Statistic of Datasets}
    \vspace{-0.25cm}
    \label{tab:datasets}
    \begin{tabular}{lcc}
      \toprule
      Dataset&JCSD&PCSD\\
      \midrule
      train & 69708 & 55538\\
      validation & 8714 & 18505\\
      test & 6489 & 18142\\
      \midrule
      Avg. tokens in code & 99.9 & 86.7\\
      Avg. tokens in comment & 17.1 & 9.4\\
      \midrule
      Max. tokens in code & 4842 & 3339\\
      Max. tokens in comment & 670 & 50\\
    \bottomrule
  \end{tabular}
  \vspace{-0.5cm}
\end{table}

We choose JCSD~\cite{hu2018summarizing} and PCSD~\cite{barone2017parallel}, which are widely used datasets for evaluating code summarization models~\cite{decom, Rencos, SGTrans}, to conduct our experiments.
JCSD contains 85K Java function-comment pairs collected from 9714 Java projects on GitHub.
Each project has at least 20 stars on GitHub. 
Given a function, JCSD regards the first sentence in its Javadoc as its summary and uses it to construct the function-comment pair.
To ensure fair comparisons, we adopt the preprocessing method used by prior work~\cite{decom, Rencos} for JCSD. 
Specifically, each function is tokenized using \textit{javalang}, and compound tokens like \textit{CamelCase} and \textit{snake\_case} are split into subtokens based on camel or snake conventions. 
Additionally, duplicate code snippets that appear in both the training and test sets are removed from the test set.
PCSD is constructed by collecting Python function-comment pairs from GitHub.
The docstring of each function is regarded as the comment.
Most prior studies preprocessed PCSD by themselves~\cite {decom, Rencos, SGTrans}. 
However, some of their preprocessing steps, such as replacing each URL link with a special token, may make different code snippets identical after preprocessing.
We find that there is an overlap between the training and test sets in each publicly available version of this dataset.
To rectify this and remove duplicate samples, we start with the dataset provided by Gao et al.~\cite{SGTrans}, use the \textit{tokenize}~\cite{tokenize} package to transform each code snippet into a sequence of lexical tokens, and remove the test samples of which the preprocessed snippets also appear in the training set.
The statistics of the two preprocessed datasets are shown in Table~\ref{tab:datasets}.

\subsection{Evaluation Metrics}
Following previous studies~\cite{decom, Rencos, editsum}, we employ five commonly used evaluation metrics: Corpus-level BLEU~\cite{bleu}, Sentence-level BLEU~\cite{bleu}, ROUGE-L~\cite{rouge}, METEOR~\cite{meteor} and CIDEr~\cite{cider}. 
BLEU calculates the accuracy based on the number of \verb|n-grams| shared between predictions and references. 
\textbf{Corpus-level BLEU} computes the geometric mean of \verb|n-grams| precisions across all predictions, assigning greater weight to longer predictions, whereas \textbf{Sentence-level BLEU} computes the arithmetic mean and assigns equal weight to each prediction.
\textbf{ROUGE-L}, a component of the ROUGE metrics, is calculated by determining the longest common subsequence between candidates and references.
According to prior work~\cite{decom, Rencos}, ROUGE-L is more suitable than other ROUGE metrics for generation tasks.
\textbf{METEOR} establishes an alignment between the generated comment and reference comment using unigrams, relying on exact matches, Porter stem matches~\cite{porter1980algorithm}, and WordNet synonymy~\cite{fellbaum1998wordnet}. 
The final score is based on the number of matching unigrams between generated comments and references.
\textbf{CIDEr} initially calculates the Sentence-level BLEU score and then adjusts it using IDF weighting. 
This adjustment assigns greater weights to infrequent words in the references but present in the prediction.

\subsection{Implementation Details}
Due to GPU memory constraints, we set the maximum token number of the code, the comment, and the input of the generator to 256, 64, and 512, respectively, and truncate the data that exceeds these limits.
We initialize our model with CodeT5, which is a state-of-the-art pre-trained encoder-decoder model for code and has been widely used by prior work~\cite{lin2023cct5, zhang2022coditt5}.
Specifically, we use the encoder of CodeT5~\cite{wang2021codet5} to initialize the neural encoder in the retriever and create a new CodeT5 instance to initialize the generator.
We use the CodeT5-base model whose parameter size is 223M.
Following the configuration of CodeT5, the dimensions of hidden states are 768, the number of heads is 12, and the number of hidden layers is 12, the learning rate is set to $5 \times 10^{-5}$ without any decay, the maximum number of epochs is set to 10 and the beam search size is set to 10.
We set k to 4 during training, i.e., retrieve the top 4 exemplars for each training sample.
The batch size is set to 24 with the gradient accumulation step size as 4. 
Training will stop early if the BLEU score on the validation set doesn't improve within two consecutive epochs. 
The experiments are conducted using four NVIDIA RTX 3090 GPUs.

\section{Results}\label{sec:Results}
We investigate the following research questions to evaluate the performance of \appname:
\begin{itemize}
    \item \textbf{RQ1}: How effective is \appname\  compared to the baselines?
    \item \textbf{RQ2}: How does the training strategy benefit \appname?
    \item \textbf{RQ3}: How does the number of exemplars used during training affect the effectiveness of \appname?
\end{itemize}
\subsection{RQ1: Comparison with Baselines}\label{sec:results:rq1}

\begin{table*}[!t]
    \vspace{-0.5cm}
    \centering
    \caption{The results of comparison with baselines}
    \vspace{-0.25cm}
    \label{tab:baseline}
    \begin{tabular}{l|ccccc|ccccc}
        \toprule
        & \multicolumn{5}{c|}{JCSD} & \multicolumn{5}{c}{PCSD} \\
        Method & C-BLEU & S-BLEU & ROUGE-L & METEOR & CIDEr &
                 C-BLEU & S-BLEU & ROUGE-L & METEOR & CIDEr \\
        \midrule
        LSI & 22.5 & 18.2 & 35.0 & 16.1 & 1.911 & 21.7 & 16.0 & 39.2 & 16.7 & 1.850 \\
        VSM & 23.3 & 19.1 & 36.6 & 17.0 & 2.038 & 23.1 & 17.1 & 41.1 & 18.0 & 2.035 \\
        NNGen & 24.2 & 19.8 & 37.5 & 17.4 & 2.092 & 24.3 & 18.1 & 42.4 & 18.6 & 2.108 \\
        \midrule
        Hybrid-DRL & - & 13.3 & 26.5 & 13.5 & 1.656 & 12.8 & 8.2 & 39.2 & 14.9 & 1.304 \\
        SG-Trans & 22.9 & 20.9 & 42.0 & 18.4 & 2.257 & 25.3 & 18.8 & 47.0 & 20.6 & 2.349 \\
        \midrule
        Rencos & - & 20.6 & 42.0 & 17.3 & 2.209 & 25.8 & 19.0 & 46.4 & 20.3 & 2.301 \\
        EditSum & - & 16.9 & 38.6 & 15.2 & 1.865 & 15.9 & 11.7 & 37.3 & 14.2 & 1.476 \\
        DECOM & 24.9 & 22.3 & 43.9 & 19.4 & 2.398 & 27.1 & 20.0 & 48.7 & 21.8 & 2.458 \\
        \midrule
        CodeT5 & 15.4 & 15.3 & 42.1 & 16.5 & 1.989 & 22.4 & 16.3 & 49.0 & 21.6 & 2.346 \\
        Unixcoder & 14.9 & 19.4 & 43.0 & 16.5 & 2.345 & 22.7 & 18.1 & 49.1 & 20.7 & 2.482 \\
        CodeT5+ & 20.3 & 19.4 & 44.3 & 18.6 & 2.332 & 26.8 & 20.3 & 51.3 & 23.3 & 2.710 \\
        \midrule
        \appname & \textbf{26.8} & \textbf{27.4} & \textbf{51.1} & \textbf{22.5} & \textbf{3.080} & \textbf{33.2} & \textbf{26.4} & \textbf{56.2} & \textbf{26.6} & \textbf{3.307} \\
        \bottomrule
    \end{tabular}
    \vspace{-0.5cm}
\end{table*}

\subsubsection{Approach} 
To evaluate the effectiveness of our proposed method, we compare our model with four kinds of comment generation models on JCSD and PCSD in terms of Corpus-level BLEU, Sentence-level BLEU, ROUGE-L, METEOR, and CIDEr.
To determine if the performance differences between the two approaches are statistically significant, we utilize the Wilcoxon signed rank test~\cite{wilcoxon1992individual}. 

\subsubsection{Baselines}

\begin{itemize}
    \item \textbf{IR-based baselines}.
    IR-based methods typically use IR techniques to retrieve similar code snippets and reuse the corresponding comments as the generated comments.
    The main difference between different works lies in the retrieval method.
    \textbf{LSI}~\cite{LSI} retrieves similar code snippets by constructing a Term Document Matrix (TDM) and reducing its dimension using Singular Value Decomposition (SVD).
    Each column in the resulting matrix is a LSI vector representing the corresponding document, and cosine similarity is used to retrieve the most similar code snippet.
    Following previous work~\cite{decom}, we set the vector dimension to be 500.
    \textbf{VSM}~\cite{VSM} is akin to LSI, but does not utilize SVD to reduce the dimension.
    Instead, VSM assigns a weight to each term in each documentation based on TF-IDF.
    \textbf{NNGen}~\cite{NNGen} represents the training and target code snippets as vectors using "bags of words". 
    Then the cosine similarity is used to narrow down the candidates, and BLEU is further employed to find the nearest neighbor of the target code. 
    
    \item \textbf{NMT-based baselines}. 
    \textbf{Hybrid-DRL}~\cite{HybridDRL} proposes to use abstract syntax trees (AST) for capturing the structural information of code and leverage the reinforcement learning framework to mitigate the discrepancy between training and testing. 
    \textbf{SG-Trans}~\cite{SGTrans} utilizes code tokens to acquire local symbolic information and the data flow graph to capture global syntactic structure. 
    It distributes the former to lower layers of transformers and the latter to higher layers, achieving outstanding performance.
    
    \item \textbf{Retrieval-Augmented approaches}. 
    These approaches combine the strengths of both NMT-based and IR-based techniques.
    For each target code snippet, \textbf{Rencos}~\cite{Rencos} first retrieves two similar code snippets based on syntactic and semantics similarities.
    It then encodes them separately and fuses them in the decoder to generate the comment. 
    Given a code snippet, \textbf{EditSum}~\cite{editsum} initiates by retrieving a similar code along with its comment using BM25.  
    It calculates the insertion and deletion sets between the two code snippets and then revises the retrieved comment based on these sets to generate the final result. 
    \textbf{DECOM}~\cite{decom} also retrieves a similar code and uses its comment as a template. 
    However, it iteratively revises the template multiple times until the comment reaches a satisfactory quality. 
    Through multiple deliberations, DECOM achieves state-of-the-art performance.
    
    \item \textbf{Pretrained models}. 
    \textbf{CodeT5}~\cite{wang2021codet5} is built upon the T5 architecture and \appname\  is initialized with the parameters of CodeT5.
    \textbf{Unixcoder}~\cite{guo2022unixcoder} is based on the transformer architecture and is designed to support both code-related understanding and generation tasks. 
    \textbf{CodeT5+}~\cite{wang2023codet5+} is pre-trained with more tasks and learns rich representation from unimodal and bimodal code-text data.
    Each pre-trained model consists of various versions with different parameter sizes.
    To ensure a fair comparison, we conduct the following experiments using CodeT5-base(223M), Unixcoder-base(126M), and CodeT5+\textsubscript{220M}.
\end{itemize}

For each baseline, if it has been trained and evaluated on the same version of JCSD or PCSD, we directly report its performance presented in its original paper.
Otherwise, we re-train and evaluate it on JCSD and/or PCSD following the settings reported in its paper.
Note that since the used PCSD is further cleaned by us, all the baselines are indeed re-trained and evaluated on our PCSD.
In addition, considering the original training set of EditSum~\cite{editsum} has 1.9M samples, which is much larger than JCSD and PCSD, we increase the maximum epoch number of EditSum from 20 to 200 to make sure that the model is fully trained and use the checkpoint which has the highest score on the validation set for evaluation.

\subsubsection{Results} 
Table~\ref{tab:baseline} lists all the results and the best performance is highlighted in bold. 
Overall, our approach \appname\  achieves the best results on the two datasets in all the metrics, followed by DECOM and CodeT5+.
On JCSD, \appname\  achieves 26.8, 27.4, 51.1, 22.5, and 3.080 points in terms of Corpus-level BLEU-4, Sentence-level BLEU-4, ROUGE-L, METEOR, and CIDEr, respectively.
Compared with the best-performing baselines, i.e., DECOM, \appname\  significantly improves the performance of the five metrics by 7.6\%, 22.9\%, 16.4\%, 16.0\%, and 28.4\%., respectively.
On the Python dataset, \appname\  achieves 33.2, 26.4, 56.2, 26.6, and 3.307 points in the five metrics. 
Compared with the best baselines, i.e., CodeT5+, \appname\  also achieves 23.9\%, 30.0\%, 9.6\%, 14.2\%, and 22.0\% improvements in the five metrics with a statistical significance, respectively.

\begin{tcolorbox}[colback=gray!10,
    colframe=black,
    width=0.49\textwidth,
    arc=2mm, auto outer arc,
    title={Answering RQ1:}, breakable]
    \appname\  outperforms the state-of-the-art baselines in all five metrics on both datasets by substantial margins. Compared to the best-performing baselines, the performance improvements of \appname\  range from 7.6\% to 30.0\%.
\end{tcolorbox}

\subsection{RQ2: Component Analysis}

\subsubsection{Training Strategy}
To demonstrate the effectiveness of our training strategy, we construct two RACG approaches which also initialize the generators with CodeT5 and keep the format of the generation input identical to \appname, but use different retrievers.

\begin{itemize}
    \item \textbf{RAF}\textsubscript{BM25} uses BM25 as the retriever.
    \item \textbf{RAF}\textsubscript{Trained Encoder} is inspired by Rencos. We first fine-tune a CodeT5 using the training set, and then use the trained encoder as the retriever.
\end{itemize}
The retrievers in both approaches work independently of their generators and are built before training the generators.
Note that, as the results on the JCSD dataset are similar to those on the PCSD dataset, we only display the results on the PCSD dataset here.

\begin{table}[!t]
    \vspace{-0.2cm}
    \centering
    \setlength{\tabcolsep}{3pt}
    \caption{Performances of different retrieval methods}
    \vspace{-0.25cm}
    \begin{tabular}{ll|ccccc}
        \toprule
        & & \multicolumn{5}{c}{PCSD} \\
        Retrieval & Generation & C-BLEU & S-BLEU & ROUGE-L & METEOR & CIDEr \\
        \midrule
        \multicolumn{2}{c|}{RAF\textsubscript{BM25}} & 31.5 & 24.4 & 54.7 & 25.8 & 3.110 \\
        \multicolumn{2}{c|}{RAF\textsubscript{Trained Encoder}} & 27.3 & 20.3 & 51.6 & 23.7 & 2.723 \\
        \midrule
        \appname & RAF\textsubscript{BM25} & 32.8 & 26.0 & 55.5 & 26.3 & 3.249 \\
        RAF\textsubscript{BM25} & \appname & 31.1 & 24.3 & 54.7 & 25.6 & 3.109 \\
        Random & RAF\textsubscript{BM25} & 11.5 & 6.3 & 41.5 & 17.1 & 1.393 \\
        Random & \appname & 13.9 & 8.7 & 44.3 & 18.3 & 1.679 \\
        \midrule
        \multicolumn{2}{c|}{\appname} & \textbf{33.2} & \textbf{26.4} & \textbf{56.2} & \textbf{26.6} & \textbf{3.307} \\
        \bottomrule
    \end{tabular}
    \label{tab:ablation}
    \vspace{-0.3cm}
\end{table}

The experimental results of these approaches are presented in Table~\ref{tab:ablation}.
We can observe that among the two approaches, \textbf{RAF}\textsubscript{BM25} achieves better performance, indicating that BM25 is a good choice when the retriever and the generator are built independently.
\appname\  significantly outperforms RAF\textsubscript{BM25} in all the metrics by substantial margins.
These results indicate our training strategy contributes to the effectiveness of our approach.

\subsubsection{Retriever and Generator}
In Section~\ref{sec:approach:training}, we hypothesize that both the retriever and the generator can benefit from our training strategy.
To verify this hypothesis, we create four variants based on \appname\  and the best RACG approach shown above, i.e., RAF\textsubscript{BM25}.
The first variant is constructed with \appname's retriever and RAF\textsubscript{BM25}'s generator.
The second variant is constructed with RAF\textsubscript{BM25}'s retriever and \appname's generator.
The third and fourth variants are constructed with a Random retriever and the generators of \appname\  and RAF\textsubscript{BM25}, respectively.
Using random sampling, the retrieved examples are mostly useless or even misleading, so this type of variant can help us evaluate the robustness of the generator.
Note that we construct each variant by connecting the corresponding retriever and generator, and do not perform additional fine-tuning.

The result is listed in Table~\ref{tab:ablation}.
\appname\  significantly outperforms the second variants, indicating the contribution of the retriever.
What's more, since RAF\textsubscript{BM25}'s generator is trained with the exemplars provided by the BM25, BM25 can be expected to fit this generator well.
However, the first variant, which uses \appname's retriever and RAF\textsubscript{BM25}'s generator, significantly outperforms RAF\textsubscript{BM25}, indicating that the exemplars retrieved by \appname\ are more useful than those retrieved by BM25 even for RAF\textsubscript{BM25}'s generator.
This further confirms the effectiveness of our retriever.
\appname\  slightly outperforms the first variant, but their performance differences are not significant.
To investigate the advantages conferred by this training strategy to the generator, we construct and compare the third and fourth variants.
We find that when the quality of exemplars is poor, the variant with \appname's generator can generate significantly better comments than the variant with RAF\textsubscript{BM25}'s generator, indicating the better robustness of \appname's generator.

\begin{tcolorbox}[colback=gray!10,
    colframe=black,
    width=0.49\textwidth,
    arc=2mm, auto outer arc,
    title={Answering RQ2:}, breakable]
    Our training strategy and the retriever contribute to the effectiveness of \appname.
    The trained retriever can be helpful even for other retrieval-augmented generators.
    The generator shows its benefit in robustness.
\end{tcolorbox}

\begin{figure*}[!t]
    \centering
    \vspace{-0.5cm}
    \begin{tikzpicture}
        \begin{axis}[
            xlabel={Number of examples on JCSD},
            xlabel style={font=\small},
            grid=major,
            width=0.4\textwidth,
            height=4cm,
            name=plot1
        ]
        \addplot coordinates{
            (2,23.46)
            (3,26.88)
            (4,26.82)
            (5,26.50)
            (6,27.04)
        };

        \addplot coordinates{
            (2,24.15)
            (3,27.63)
            (4,27.43)
            (5,27.35)
            (6,27.42)
        };

        \addplot coordinates{
            (2,20.75)
            (3,22.55)
            (4,22.48)
            (5,22.32)
            (6,22.54)
        };
        \end{axis}
    \end{tikzpicture}
    \begin{tikzpicture}
        \begin{axis}[
            xlabel={Number of examples on PCSD},
            xlabel style={font=\small},
            grid=major,
            width=0.4\textwidth,
            height=4cm,
            at=(plot1.east),
            anchor=west,
            name=plot2,
            legend pos=outer north east,
        ]

        \addplot coordinates{
            (2,27.58)
            (3,27.04)
            (4,33.19)
            (5,32.97)
            (6,33.13)
        };
        \addlegendentry{C-BLEU4}

        \addplot coordinates{
            (2,20.83)
            (3,20.36)
            (4,26.41)
            (5,26.09)
            (6,26.41)
        };
        \addlegendentry{S-BLEU4}

        \addplot coordinates{
            (2,23.70)
            (3,23.43)
            (4,26.62)
            (5,26.49)
            (6,26.47)
        };
        \addlegendentry{METEOR}
            
        \end{axis}
    \end{tikzpicture}
    \vspace{-0.3cm}
    \caption{Performance on different number of examples}
    \label{fig:number}
\end{figure*}
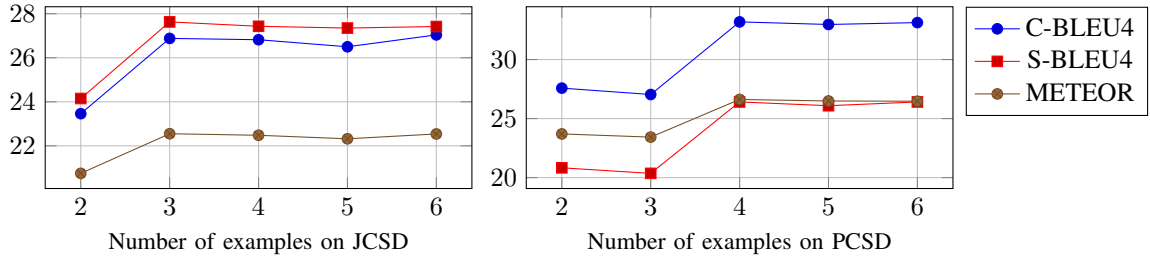

\subsection{RQ3: Analysis of the Exemplar Number}

During training, \appname\  retrieves top-$k$ exemplars to calculate the loss and jointly train the retriever and the generator.
The number of the retrieved exemplars, i.e., $k$, may affect the performance of \appname.
In addition, the larger the $k$ is, the longer the training time will be.
To find a $k$ that can balance time and performance, we train multiple \appname\  models with different $k$s.
If only one exemplar is retrieved, i.e., $k=1$, the weight of the exemplar will always be 1(c.f. Equation~\ref{eq:topk_prob}), making the retriever unable to learn from generation.
So the minimal value of $k$ is set to two.
The maximum value of $k$ is set to six due to time and resource constraints.
Considering that the value ranges for ROUGE-L and CIDEr differ significantly from the other three metrics and our method exhibits similar performance trends across all the metrics, we only display the results for Corpus-BLEU, Sentence-BLEU, and METEOR.
Please note that $k$ is used only during training.

Figure~\ref{fig:number} presents the results. 
On JCSD, the performance of JointCom increases when k goes from 2 to 3 and remains similar when k is between 3 and 6.
On PCSD, the performance of \appname\  first remains similar when $k$ is enlarged from 2 to 3 and then increases when $k$ goes from 3 and 4.
Thus, we consider $k=4$ to be a trade-off choice between time and performance.
We also notice that when $k$ is 2, \appname's performance is worse than that of RAF\textsubscript{BM25} which indicates that a small exemplar number may lead to insufficient training.

\begin{tcolorbox}[colback=gray!10,
    colframe=black,
    width=0.49\textwidth,
    arc=2mm, auto outer arc,
    title={Answering RQ3:}, breakable]
    Overall, the model converges after 4.
    Thus we consider $K=4$ to be a trade-off choice between effectiveness and efficiency.
\end{tcolorbox}

\section{Human Evaluation}\label{sec:HumanEvaluation}
Although the evaluation metrics, e.g., BLEU, ROUGE-L, and METEOR, can reflect the lexical similarities between the generated comments and the references, these metrics are not always consistent with human evaluations. Because they do not necessarily capture the fluency, naturalness, and other subjective factors related to comments~\cite{callison2006re}.
Hence, we conduct a human evaluation to further assess the quality of comments generated by various approaches.

\begin{table*}[!t]
    \centering
    \begin{threeparttable}
    \caption{The results of human evaluation}
    \label{tab:question}
    \vspace{-0.25cm}
    \begin{tabular}{c|ccc|ccc}
        \toprule
        \multirow{2}{*}{Approach} & \multicolumn{3}{c|}{JCSD} & \multicolumn{3}{c}{PCSD} \\
                                    & Naturalness & Informativeness & Usefulness & Naturalness & Informativeness & Usefulness \\
        \midrule
        CodeT5+ & $4.52 \pm 0.76(0.189)$ & $3.52 \pm 0.83(\textless 0.05)$ & $3.74 \pm 1.00(\textless 0.05) $ & $4.36 \pm 0.94(0.295)$ & $3.52 \pm 0.88(\textless 0.05)$ & $3.61 \pm 1.08(\textless 0.05)$ \\
        DECOM & $4.59 \pm 0.71(0.409)$ & $3.55 \pm 0.98(\textless 0.05)$ & $3.71 \pm 1.04(\textless 0.05)$ & $4.52 \pm 0.78(0.831)$ & $3.53 \pm 0.97(\textless 0.05)$ & $3.54 \pm 1.15(\textless 0.05)$ \\
        \appname & \bm{$4.63 \pm 0.68$} & \bm{$3.76 \pm 0.87$} & \bm{$3.95 \pm 0.99$} & \bm{$4.56 \pm 0.85$} & \bm{$3.82 \pm 0.85$} & \bm{$3.91 \pm 1.06$} \\
        \bottomrule
    \end{tabular}
    \begin{tablenotes}[center]
        \footnotesize
        \item[*] The number in parentheses is the p-value in the Wilcoxon signed-rank test.
    \end{tablenotes}
    \end{threeparttable}
    \vspace{-0.5cm}
\end{table*}

\subsection{Procedure}
We recruit eight participants, including four Ph.D. students and four master students, who are not co-authors of this paper. 
They all have at least three years of both Java and Python development experience, and four of them have more than five years of development experience. 
We randomly select 100 code snippets from the test sets (50 from JCSD and 50 from PCSD).
Employing \appname\  and the two baselines with the best performance, i.e., Decom and CodeT5+, we generate a total of 300 comments.

We divide the 100 samples into four groups and convert each group into a questionnaire.
Following prior work~\cite{decom, re2com, editsum}, each participant is asked to rate each generated comment from three aspects: 
(1) Naturalness, reflecting the fluency of the generated comments in terms of grammar; 
(2) Informativeness, reflecting the richness of information in the generated comments;
and (3) Usefulness, reflecting how helpful the generated comments can be for developers.
All three scores are integers, ranging from 1 to 5 (1 for poor and 5 for excellent).
The explanation of each aspect as well as two examples are presented in the header of each questionnaire.
For each sample, its code snippet is presented first.
To help participants correctly understand the code, we attach the ground truth after the code snippets.
Then the comments generated by different approaches for this code snippet are randomly listed for the participants to assess.
The approach names are removed to ensure that the participants are not aware of which approaches the comments are generated by. 
Each questionnaire is evaluated by two participants, and the final score for a generated comment is the average of the two participants' ratings.

\subsection{Results}

Table~\ref{tab:question} exhibits the results of our human evaluation.
Overall, \appname\  outperforms both CodeT5+ and DECOM on the two datasets in all aspects, which is in line with our results in Section~\ref{sec:results:rq1}.
On JCSD, \appname\  achieves average scores of 4.63, 3.76, and 3.95 in terms of naturalness, informativeness, and usefulness, respectively.
On PCSD, such scores are 4.56, 3.82 and 3.91.
Specifically, although our approach only slightly outperforms baselines in terms of naturalness, the average scores of \appname\  and baselines are both over 4.5, indicating that \appname\  can generate fluent and natural comments.
In terms of informativeness and usefulness, \appname\  improves over the best-performing baselines by 0.250 and 0.255 points on average.
We also use the Wilcoxon signed-rank test to assess the significance of performance differences between JointCom and the baselines. 
The results confirm that the comments generated by our approach are more informative and useful than the baselines for developers.
Although the difference in the naturalness aspect is not statistically significant, we argue it would be safe to conclude that \appname\  demonstrates superiority over CodeT5+ and DECOM.

\section{Discussion}\label{sec:Discussion}
\subsection{Qualitative Analysis}

\begin{table}[!t]
    \vspace{-0.2cm}
    \small
    \centering
    \caption{Example of qualitative analysis-1}
    \label{tab:example1}
    \vspace{-0.25cm}
    \begin{threeparttable}
    \addtolength\tabcolsep{2.8pt}
    \begin{tabular}{|p{0.47\textwidth}@{\hskip3pt}|}
         \toprule
         \textbf{Code}:\\
         \vspace{-0.5cm}
         \begin{lstlisting}[language=Python]
def _build_status(data, item): 
    stream = item['stream'] 
    if ('Running in' in stream): 
        data.setdefault('Intermediate_Containers', []).append(stream.rstrip().split()[(-1)]) 
    if ('Successfully built' in stream): 
        data['Id'] = stream.rstrip().split()[(-1)]
         \end{lstlisting}\vspace{-0.4cm}\\
         \textbf{Comment}:\\
         \vspace{-0.3cm}
         \textcolor{blue}{process a status update from a docker build}. \\
         \hline
         \textbf{Retrieved comment}\\
         \textit{RAF}\textsubscript{BM25}: serialize a list of qwebhistoryitems to a data stream.\\
         \textit{RAF}\textsubscript{Trained Encoder}: print only the keys for an item.\\
         \textit{DECOM}: serialize a list of qwebhistoryitems to a data stream.\\
         \textit{\appname}: \textcolor{blue}{process a status update from a docker} push.\\
         \hline
         \textbf{Prediction}\\
         \textit{RAF}\textsubscript{BM25}: \textcolor{blue}{build} the \textcolor{blue}{status} for a job.\\
         \textit{RAF}\textsubscript{Trained Encoder}: check \textcolor{blue}{status} of \textcolor{blue}{build process}.\\
         \textit{DECOM}: \textcolor{blue}{build} a \textcolor{blue}{status} item from a stream.\\
         \textit{\appname}: \textcolor{blue}{process a status update from a docker build}.\\
         \bottomrule
    \end{tabular}
    
    \end{threeparttable}
    \vspace{-0.5cm}
\end{table}

\begin{table}[!t]
    \small
    \centering
    \caption{Example of qualitative analysis-2}
    \label{tab:example2}
    \vspace{-0.25cm}
    \begin{threeparttable}
    \addtolength\tabcolsep{2.8pt}
    \begin{tabular}{|p{0.47\textwidth}@{\hskip3pt}|}
         \toprule
         \textbf{Code}:\\
         \vspace{-0.5cm}
         \begin{lstlisting}[language=Python]
def dmp_neg(f, u, K): 
    if (not u): 
        return dup_neg(f, K) 
    v = (u - 1) 
    return [dmp_neg(cf, v, K) for cf in f]
         \end{lstlisting}\vspace{-0.4cm}\\
         \textbf{Comment}:\\
         \vspace{-0.3cm}
         \textcolor{blue}{negate a polynomial in k[x]}. \\
         \hline
         \textbf{Retrieved comment}\\
         \textit{RAF}\textsubscript{BM25}: quotient by a constant \textcolor{blue}{in k[x]}.\\
         \textit{RAF}\textsubscript{Trained Encoder}: multiply f by a constant value \textcolor{blue}{in k[x]}.\\
         \textit{DECOM}: quotient by a constant \textcolor{blue}{in k[x]}.\\
         \textit{\appname}: exact quotient by a constant \textcolor{blue}{in k[x]}.\\
         \hline
         \textbf{Prediction}\\
         \textit{RAF}\textsubscript{BM25}: \textcolor{blue}{negative} by a constant \textcolor{blue}{in k[x]}. \\
         \textit{RAF}\textsubscript{Train Encoder}: \textcolor{blue}{negates }f by a constant value \textcolor{blue}{in k[x]}. \\
         \textit{DECOM}: multiply f by a constant value \textcolor{blue}{in k[x]}.\\
         \textit{\appname}: \textcolor{blue}{negate a polynomial in k[x]}.\\
         \bottomrule
    \end{tabular}
    
    \end{threeparttable}
    \vspace{-0.5cm}
\end{table}

\begin{table}[!t]
    \vspace{-0.2cm}
    \small
    \centering
    \caption{Example of qualitative analysis-3}
    \label{tab:example_bad}
    \vspace{-0.25cm}
    \begin{threeparttable}
    \addtolength\tabcolsep{2.8pt}
    \begin{tabular}{|p{0.47\textwidth}@{\hskip3pt}|}
         \toprule
         \textbf{Code}:\\
         \vspace{-0.5cm}
         \begin{lstlisting}[language=Java]
static String toLowerCase(String s) {
  int len = s.length();
  StringBuilder sb = null;
  for (int i = 0; i < len; i++) {
    char c = s.charAt(i);
    if (c >= 'a' && c <= 'z' || c == '.') {
      if (sb != null) sb.append(c);
    } else if (c >= '0' && c <= '9' || c == '-') {
      if (sb != null) sb.append(c);
    } else if (c >= 'A' && c <= 'Z') {
      if (sb == null) {
        sb = new StringBuilder(len);
        sb.append(s, 0, i);
      }
      sb.append((char)(c - CASE_DIFF));
    } else {
      throw new IllegalArgumentException("Invalid characters in hostname");
    }
  }
  return sb == null ? s : sb.toString();
}
         \end{lstlisting}\vspace{-0.4cm}\\
         \textbf{Comment}:\\
         \vspace{-0.3cm}
         \textcolor{blue}{convert to lower case, and check that all chars are ascii alphanumeric, '-' or '.' only}. \\
         \hline
         \textbf{Retrieved comment}\\
         \textit{\appname}: \textcolor{blue}{converts} all of the \textcolor{blue}{characters} in the string to upper \textcolor{blue}{case}, based on the locale.\\
         \hline
         \textbf{Prediction}\\
         \textit{\appname}: \textcolor{blue}{converts all} of the \textcolor{blue}{characters} in the string \textcolor{blue}{to lowercase}.\\
         \bottomrule
    \end{tabular}
    
    \end{threeparttable}
    \vspace{-0.3cm}
\end{table}

For qualitative analysis of our approach, we present two cases to elucidate \appname's efficacy.
Each case presents the target code snippet, the target comment, the comments retrieved by different retrievers, and the comments generated by different approaches.
In case 1 shown in Table~\ref{tab:example1}, \appname\  retrieves an excellent comment which is the same as the target comment except the object, and the generator can generate the target comment by simple modification. 
However, BM25 and the retriever of RAF\textsubscript{Trained Encoder} retrieve some unrelated information which misleads the generators to predict incorrect verbs. 
In case 2 shown in Table~\ref{tab:example2}, \appname, BM25, and the retriever of RAF\textsubscript{Trained Encoder} all retrieve comments containing the wrong action.
In this case, \appname\  can overlook the useless part in the exemplar and generate a good comment. 
However, the generators of RAF\textsubscript{BM25} and RAF\textsubscript{Trained Encoder} are negatively affected by the retrieved information and produce bad results.

According to these examples, we believe that \appname\  performs better than the baselines for two main reasons. 
First, the retriever of \appname\  is more suitable for comment generation after being jointly trained with the generator and can retrieve better exemplars to assist the generator. 
Second, the generator of \appname\  is more robust and can generate appropriate comments even if the exemplars are inappropriate.

We also present a case where \appname\  underperforms, as shown in Table~\ref{tab:example_bad}.
In this case, the primary tasks of the input code involve converting the input string to lowercase and verifying that all characters are ASCII alphanumeric, '-' or '.' only.
\appname\  recognizes the conversion task but fails to encapsulate the verification aspect.
This case underscores the challenge \appname\  faces in delving deeper into the nuanced details of a function, which needs further study.

\subsection{Feasibility on LLMs}
\begin{table}[!t]
    \vspace{-0.2cm}
    \centering
    \setlength{\tabcolsep}{3pt}
    \caption{Enhancing LLMs with Different Retrieval Methods}
    \vspace{-0.25cm}
    \begin{tabular}{c|ccccc}
        \toprule
        & \multicolumn{5}{c}{PCSD} \\
        Retrieval & C-BLEU & S-BLEU & ROUGE-L & METEOR & CIDEr \\
        \midrule
        BM25 & 6.0 & 5.8 & 27.5 & 18.6 & 0.819 \\
        encoder\textsubscript{trained} & \textbf{11.1} & \textbf{10.4} & \textbf{37.6} & \textbf{21.1} & \textbf{1.372} \\
        \bottomrule
    \end{tabular}
    \label{tab:llm}
    \vspace{-0.6cm}
\end{table}

Currently, large language models(LLMs) have achieved remarkable results on various code-related tasks. 
To demonstrate the feasibility of our proposed framework on large language models, we evaluate CodeLlama-7b-Instruct~\cite{roziere2023code}, which is widely used by previous work for code-related tasks~\cite{liu2024your}, on the PCSD dataset with in-context learning.
For retrieval, we employ BM25 and an encoder trained with our framework to retrieve an exemplar for each input code.
Restricted by our computing resources, we freeze the parameters of the LLM and only train the retriever based on the LLM's output.
For generation, we prompt the LLM to generate a one-sentence comment for each input code with the retrieved exemplar as the demonstration.
The results are shown in Table~\ref{tab:llm}.
The encoder trained with our framework outperforms BM25 in terms of all the metrics by substantial margins, indicating that it can retrieve better exemplars for the LLM.
These results demonstrate that our framework can benefit LLMs without finetuning it.

\subsection{Threats to Validity}
The first threat to validity is that the datasets we employ only include Java and Python code. 
However, Java and Python are two of the most popular programming languages. 
Also, our approach is language-agnostic and can be seamlessly applied to other programming languages.

The second threat relates to the evaluation metrics utilized in this experiment.
Existing automatic evaluation methods often fall short of maintaining complete consistency with human preferences, as they may focus on specific aspects while neglecting other dimensions~\cite{callison2006re, gros2020code}. 
To mitigate this limitation, we use five evaluation metrics to compare \appname\  with the baselines. 
There are also some threats related to our human evaluation.
Firstly, we cannot guarantee that every participant has no subjective bias and fairly evaluates each comment.
To mitigate this threat, each comment is evaluated by two participants and we use the average score as the final score.
Secondly, the potential for incorrect ground truths poses a challenge, as it may mislead participants and yield scores that do not reflect the quality of the generated comments. 
To investigate this threat, we randomly sampled 50 code-comment pairs from the test set and manually verified the correctness of each ground truth comment.
We found that 47 out of 50 of the comments are correct.
Thus, we believe this threat is limited.

The third concern pertains to potential errors in the implementation of baselines.
To mitigate this threat, we directly use the publicly available code of the baselines to conduct experiments.
However, EditSum did not provide the scripts for preprocessing data, so we tried our best to re-implement this part carefully according to the corresponding paper.

The fourth threat lies in the approximation in our training stage.
If the best exemplar is not among the k retrieved exemplars, the retriever will be trained to retrieve a suboptimal exemplar.
To alleviate this problem, we initialize the retriever with the parameters of CodeT5's encoder to ensure the quality of the top-k exemplars.
It is worth mentioning that the search index $M$ is updated after each epoch, which can further help the retriever to find the best exemplars during training.

\section{Related Work}\label{sec:RelatedWork}
\subsection{Automatic Comment Generation}
Comment generation has become a rapidly developing research topic for years.
Early studies focused on leveraging manually crafted templates to generate comments~\cite{moreno2013automatic, mcburney2015automatic, sridhara2011automatically, sridhara2010towards}.
These methods require researchers to design a set of templates, then automatically extract some keywords from the source code and fill keywords into a suitable template.
For example, Sridhara et al.~\cite{sridhara2010towards} proposed an approach to analyze the code snippet to identify the action, theme, and secondary arguments, and then generate the natural language description.
However, it's hard for the predefined templates to cover all situations and it is also possible that there is no useful keyword due to non-standard identifier naming.
Some researchers used IR techniques to extract keywords from the source code and compose them into term-based comments~\cite{haiduc2010supporting, haiduc2010use, li2017summary, eddy2013evaluating}.
For example, Haiduc et al.~\cite{haiduc2010use} used VSM~\cite{VSM} and LSI~\cite{LSI} methods to encode code snippets and code tokens into vectors, and selected the keywords according to the cosine similarity between code and tokens. 
Other researchers used IR techniques to detect code clones and reuse the comment of the code clone as the generated code comment~\cite{wong2013autocomment, wong2015clocom, NNGen}. 
However, the former suffers from non-standard identifier naming like template-based method and the performance of the latter greatly depends on the quality of the code retrieval base. 
In recent years, deep-learning-based methods become more and more popular and achieved remarkable results~\cite{CodeNN, hu2018deep, re2com, decom, SGTrans, Rencos, chen2018neural, leclair2021ensemble}. 
Iyer et al.~\cite{CodeNN} proposed CODE-NN which used LSTM and the attention mechanism to "translate" code to comments. 
To help neural networks better generate low-frequency words, researchers proposed Retrieval-Augmented Comment Generation (RACG) approaches, which combine IR-based methods with neural network-based methods.
Wei et al.~\cite{re2com} proposed Re2Com. 
Given a code snippet, Re2Com retrieved a similar code snippet from the retrieval base, used the corresponding comment as part of the input, and adjusted the weights of the code snippet and the comment according to the cosine similarity between the target code and the retrieved code.

Our approach is also a RACG approach.
Different from existing RACG approaches, our approach uses a training strategy to jointly train the retriever and the generator, which results in a better retriever for comment generation and a more robust generator.
The experimental results also demonstrate the superiority of our approach.

\subsection{Code Clone Detection and Code Retrieval}
The retriever in our approach is also related to code clone detection and code retrieval.
Code clone detection, which aims to find the duplication of source code, is an active area of research.
Early studies focus on token-based methods~\cite{roy2009comparison, jadon2016code, nakamura2016towards}.
For example, Roy et al.~\cite{roy2009comparison} proposed NiCad, a token-based clone detection tool that identifies both exact clones and renamed clones.
Some researchers detect code clones based on AST matching~\cite{yang2018structural, pati2017comparison, chodarev2015haskell}, which is good at dealing with the addition or removal of statements.
For example, Chodarev et al.~\cite{chodarev2015haskell} presents an algorithm for code clone detection based on comparing parts of ASTs and finding repeating patterns.
With the development of deep learning, neural networks have been applied to detect code clones~\cite{allamanis2017learning, white2016deep}.
For instance, White et al.~\cite{white2016deep} proposed a learning-based detection technique that leveraged both lexical- and syntactic-level patterns.

Code retrieval aims to search code snippets according to natural language descriptions.
Current methods can be divided into two categories: information-retrieval-based methods~\cite{haiduc2013automatic, hill2014nl, lu2015query} and deep-learning-based methods~\cite{allamanis2015bimodal, gu2018deep}.
For example, To address the problem that the words used in a query may be different from the words that have similar semantics in the source code, Lu et al.~\cite{lu2015query} used synonyms generated by WordNet to extend the query.
Gu~\cite{gu2018deep} used RNN to encode the query and the code snippets into feature vectors and then ranked the candidate set according to cosine similarities.

Different from these studies, our approach targets code comment generation, and its retriever aims to retrieve code-comment pairs that can help generate the target comment.
In addition, we improve the retriever for code comment generation by jointly training it with the generator.

\section{Conclusion}\label{sec:Conclusion}

In this paper, we propose to improve retrieval-augmented comment generation (RACG) by jointly training retrievers and generators.
To achieve this goal, we design a new training strategy and implement a RACG approach named \appname\  based on it.
Compared to existing RACG approaches, the retriever of \appname\  can retrieve exemplars that are useful for generation and the generator is more robust.
The evaluation results show that our approach outperforms the state-of-the-art baselines on both Java and Python datasets.
A human evaluation also confirms that the comments generated by \appname\  are more natural, informative, and useful.
In the future, we plan to investigate the applicability of this framework to other code-related tasks, such as bug fixing and code translation, and to pre-trained models other than CodeT5, such as CodeT5+.
We have released our replication package, including the used datasets and our source code, at \url{https://github.com/HanzhenLu/JointCom}.

\section*{Acknowledgments}
This research/project is supported by the National Natural Science Foundation of China (No. 62202420) and the Fundamental Research Funds for the Central Universities (No. 226-2022-00064). Zhongxin Liu gratefully acknowledges the support of Zhejiang University Education Foundation Qizhen Scholar Foundation.

\balance
\bibliographystyle{IEEEtran}
\bibliography{cite}

\end{document}